\documentclass[preprint,12pt, a4paper]{elsarticle}

\usepackage{amssymb}
\usepackage{hyperref}
\usepackage{booktabs}
\usepackage{listings}
\usepackage{float}
\usepackage{placeins}
\usepackage{tikz}
\usetikzlibrary{positioning,arrows.meta}

\newsavebox{\shotbox}

\newcommand{\apexversion}{0.1.1}
\newcommand{\apexpython}{3.12}
\newcommand{\apexnode}{22.12}
\newcommand{\apexlayers}{eight}
\newcommand{\apexregistries}{31}
\newcommand{\apexconfigregistries}{28}
\newcommand{\apexapprovalruns}{20}
\newcommand{\apexapprovalsimruns}{20,000}
\newcommand{\apexparitytol}{10^{-12}}
\newcommand{\apexpymootol}{10^{-13}}
\newcommand{\apextestspassed}{3,129}
\newcommand{\apextestsfailed}{no}
\newcommand{\apextestsskipped}{40}
\newcommand{\apextestsfailnote}{; of the skipped tests, 35 need optional extras, 4 are opt-in tests with language models and 1 has an empty parameter set}
\newcommand{\apexuitests}{107}

\newcommand{\apexcollapse}{5\%}
\newcommand{\apexpymoohvtol}{10^{-12}}
\newcommand{\apexpopttol}{10^{-8}}

\journal{SoftwareX}

\begin{document}
\renewcommand{\labelenumii}{\arabic{enumi}.\arabic{enumii}}

\begin{frontmatter}

\title{APEX: Agentic Platform for Ensemble eXploration}

\author[norce]{Xiaodong Luo\corref{cor1}}
\ead{xluo@norceresearch.no}
\cortext[cor1]{Corresponding author}
\address[norce]{NORCE Norwegian Research Centre, P.O.B 22 Nyg{\aa}rdstangen, NO-5838 Bergen, Norway}

\begin{abstract}
Ensemble-based methods provide a practical framework for history matching, prediction, and optimization of subsurface models. In methodological studies, however, a method is seldom assessed through a single run. Comparisons normally span cases, configurations, and random seeds, while each evaluation may require an expensive external simulator. Keeping such studies computationally bounded, auditable, and reproducible therefore requires more than an implementation of the numerical method: experiments must be validated before execution, their computational costs must be estimated, repeated runs must be organized consistently, and their configurations and outputs must be traceable. APEX (Agentic Platform for Ensemble eXploration), an open-source Python platform, was thus developed to address these requirements, with a common framework for configuring, validating, executing, recording, and comparing ensemble-based experiments.

APEX organizes its functionality around three related concepts: an experiment defines what is to be executed, a run records one execution of that experiment, and a study coordinates and compares multiple experiments or runs. An experiment is specified by a validated configuration, which can also be generated by a planner agent. Before execution, APEX estimates the maximum computational cost in terms of forward simulations. If this estimate exceeds a preset threshold, submission of the experiment is held for user approval. Each run records provenance, whereas a study repeats a problem across variants and paired replicates and compares the variants using paired statistics. The scientific and computational components used in these workflows, including methods, simulators, metrics, and optimizers, are provided as plugins. A command-line interface, a Python client, a Representational State Transfer (REST) service with a web interface, and tools for language-model agents access these capabilities through the same service layer.

APEX is licensed under GPL-3.0-or-later; release v\apexversion{} is available at \url{https://github.com/lanhill/APEX-public}.
\end{abstract}

\begin{keyword}
Agentic ensemble workflows \sep History matching \sep Production optimization \sep Prediction \sep Python \sep Reproducibility
\end{keyword}

\end{frontmatter}

\section*{Required Metadata}
\label{sec:metadata}

\section*{Current code version}
\label{sec:codeversion}

\begin{table}[H]
\begin{tabular}{|l|p{6.5cm}|p{6.5cm}|}
\hline
\textbf{Nr.} & \textbf{Code metadata description} & \textbf{Please fill in this column} \\
\hline
C1 & Current code version & v\apexversion{} (release tag; package version \apexversion) \\
\hline
C2 & Permanent GitHub link to code/repository used for this code version & \url{https://github.com/lanhill/APEX-public/tree/v0.1.1}; archived on Zenodo, \href{https://doi.org/10.5281/zenodo.23780462}{doi:10.5281/zenodo.23780462} \\
\hline
C3 & Legal Code License   & GPL-3.0-or-later; parts adapted from pymoo carry its Apache-2.0 licence notice \\
\hline
C4 & Code versioning system used & git \\
\hline
C5 & Software code languages, tools, and services used & Python; TOML (configurations); TypeScript and React (web interface) \\
\hline
C6 & Compilation requirements, operating environments \& dependencies & Linux; Python $\geq$ \apexpython, installed with uv from a locked environment; NumPy, SciPy, pydantic, PyWavelets, Typer; optional extras (e.g., FastAPI for the REST service, MCP, matplotlib); OPM Flow for reservoir simulation; Eclipse only where installed and licensed; Node.js $\geq$ \apexnode{} to build the web interface \\
\hline
C7 & If available Link to developer documentation/manual & Documentation in the repository's \texttt{docs/} folder, \url{https://lanhill.github.io/APEX-public/}, with six getting-started pages, and narrated tutorial videos with subtitles in the repository's \texttt{videos/} folder and on YouTube (playlist ``APEX tutorials'', \url{https://www.youtube.com/playlist?list=PLLxKDH7QwDR4}) \\
\hline
C8 & Support email for questions & xluo@norceresearch.no \\
\hline
\end{tabular}
\caption{Code metadata (mandatory)}
\label{tab:metadata}
\end{table}

\section{Motivation and significance}
\label{sec:motivation}

Ensemble-based history matching methods, such as the ensemble smoother (ES) \cite{skjervheim2011ensemble,van1eeuwen996data} and their iterative variants \cite{chen2013-levenberg,emerick2012ensemble, luo2015Iterative,evensen2019efficient}, are widely used to estimate uncertain parameters of subsurface models from observed data. These methods are derivative-free and non-intrusive with respect to the forward simulator, can handle high-dimensional parameterizations, and represent uncertainty to some extent through an ensemble of reservoir models \cite{Oliver2008}. The resulting ensembles can subsequently support production prediction and optimization under geological and operational uncertainty.

Methodological research ensemble-based methods seldom ends with a single run. A localization scheme \cite{luo2018correlation_spej,luo2019automatic}, a hyper-parameter optimization procedure \cite{luo2022continuous,luo2023hyper}, a stopping criterion \cite{luo2024better}, or a strategy for imposing soft constraints \cite{luo2022data} is normally examined against alternatives across cases and configurations. Such comparisons are particularly demanding when one evaluation requires many calls to an external reservoir simulator. The scientific conclusion may then depend not only on the numerical method, but also on how experiments are configured, paired, executed, resumed and recorded.

In practice, four issues recur. First, the number of simulator runs required by an experiment may not be made explicit before execution, even though simulator time, processor capacity or software licences are limited. Second, a comparison based on a single run per method, or on runs that do not share the same random draws, may confound methodological differences with sampling variability. Third, the configuration, inputs, software revision and execution environment behind a reported quantity may be incomplete, which weakens the auditability and reproducibility of reported results. Fourth, numerical methods and supporting components are distributed across software packages with different interfaces, so a controlled comparison may require substantial engineering in addition to the scientific implementation.

For subsurface problems, several open-source packages address parts of this workflow. PET \cite{PETSoftware}, developed at Norwegian Research Centre (NORCE), provides ensemble-based data assimilation and optimization and is the source from which APEX (Agentic Platform for Ensemble eXploration) ports a number of methods. 
ERT \cite{ERTSoftware} provides ensemble-based history matching workflows for reservoir models, while Everest \cite{Hanea2025Everest} supports decision making under uncertainty. PEST++ \cite{White2020PESTpp} provides a suite for model calibration with an iterative ensemble smoother (IES). DAPPER \cite{Raanes2024DAPPER} supports controlled comparisons of data-assimilation methods over experiments and random seeds, primarily targeting benchmark models with filtering and smoothing data assimilation algorithms. General experiment management software, such as Sacred \cite{Greff2017Sacred}, records configurations and provenance. These packages differ in scientific scope, simulator coupling and experiment management objectives; they are therefore complementary rather than direct substitutes.

Within this context, the unresolved software problem considered here is how to place ensemble-based experiments with expensive external simulators in one workflow that combines pre-execution cost planning, paired comparisons over random seeds, provenance, check-pointing and access through several user interfaces. This gap is stated in terms of combined requirements, not as an absence of capable methods or workflow systems in individual areas. APEX \cite{APEXSoftware} is thus developed to address this combined workflow problem by providing a common framework for planning experiments, recording and resuming individual runs, and organizing paired comparisons across configurations and random seeds.

APEX is developed for ensemble-based history matching, forecasting, and optimization of subsurface models. It re-implements selected methods and components from some third-party repositories like PET, ERT, Everest, and pymoo \cite{BlankDeb2020Pymoo} and integrates them into respective agentic workflows, while providing additional functionality within a common plugin architecture. APEX also converts PET (main branch commit e65fb10) and ERT (version 26.0.1) configurations and checks ported implementations against stored reference outputs when suitable reference data are available (Section~\ref{sec:quality}). Each run uses explicit random number generators whose seeds are defined in the configuration and stored in check points. Because APEX maintains its own implementations, updates from the source repositories must be separately ported, tested, and validated before becoming available in APEX.

In APEX, a user specifies a problem in a TOML configuration file, validates the configuration to obtain a run plan and its cost bound, submits the run, and reads the posterior summary through the command line, Python client or web interface. A study file expands the same problem over cases, methods, variants and random seeds. 
The name APEX also refers to runtime agents and tools for language-model agents (Section~\ref{sec:functionalities}); however, the experiment management workflow presented in this paper can also work without an agent. 
The agent-facing components in APEX are provided as interfaces and advisory functions for more automated workflow initialization, execution, monitoring, diagnosis and report, not as validated scientific decision makers.

In short, APEX possesses the following features. First, an experiment is represented by a validated configuration whose maximum simulator cost is planned before execution, or is held for user approval when the cost exceeds a preset bound. 
Second, runs are identified by their content, check-pointed and recorded with provenance. 
Third, studies compare variants over paired seeds using paired differences and Wilcoxon signed-rank tests \cite{wilcoxon1945individual}. 
Fourth, methods ported from a third-party repository are checked against reference outputs. 
Fifth, the same validation and execution functionalities are accessible through interactive interfaces (web UI) for users and programmatic interfaces for scripts and language-model agents.

The remainder of the paper is organized as follows. Section~\ref{sec:description} describes the architecture, functionalities and software-quality checks; Section~\ref{sec:examples} illustrates the workflow; Section~\ref{sec:impact} discusses its potential impact and limitations; and Section~\ref{sec:conclusions} summarizes the work.

\section{Software description}
\label{sec:description}

\subsection{Software architecture}
\label{sec:architecture}

\begin{figure}[t]
\centering
\resizebox{\linewidth}{!}{%
\begin{tikzpicture}[
  layer/.style={draw, rounded corners=2pt, text width=9.8cm, minimum height=0.9cm,
    align=center, font=\footnotesize},
  side/.style={draw, dashed, rounded corners=2pt, text width=2.8cm, minimum height=0.9cm,
    align=center, font=\footnotesize},
  node distance=1.6mm]
\node[layer, fill=blue!8] (fe) {\textbf{Front ends}: command line $\cdot$ Python client (in-process or remote)\\ REST service with web interface $\cdot$ tools for language-model agents (MCP)};
\node[layer, below=of fe] (int) {\textbf{Integrations}: ERT and Everest (optional)};
\node[layer, below=of int] (agt) {\textbf{Agents}: monitor, diagnosis, planner, report, plotter\\ case-converter, geomodeler};
\node[layer, below=of agt, fill=blue!4] (svc) {\textbf{Services}: validation and run plans $\cdot$ job service and approval\\ studies and comparison $\cdot$ run identity, reuse and provenance};
\node[layer, below=of svc] (wf) {\textbf{Workflows}: history matching $\cdot$ prediction $\cdot$ optimization};
\node[layer, below=of wf] (alg) {\textbf{Algorithms} $\cdot$ \textbf{Simulation} $\cdot$ \textbf{Execution}\\ \textbf{Storage} $\cdot$ \textbf{Reporting}};
\node[layer, below=of alg] (core) {\textbf{Core}: protocols and plugin registries};
\node[layer, below=of core] (dom) {\textbf{Domain}: configuration schemas (strict)};
\node[side, anchor=west] (user) at ([xshift=8mm]fe.east) {Users, scripts,\\external agents};
\node[side, anchor=west] (sim) at ([xshift=8mm]alg.east) {External simulators\\(OPM Flow, Eclipse)};
\node[side, anchor=west] (plug) at ([xshift=8mm]dom.east) {Plugin packages\\(Python\\entry points)};
\draw[{Stealth}-{Stealth}] (user) -- (fe);
\draw[{Stealth}-{Stealth}] (sim) -- (alg);
\draw[-{Stealth}] (plug.north west) -- (core.east);
\draw[-{Stealth}, thick] ([xshift=-5mm]fe.north west) -- node[left, font=\footnotesize, align=center] {imports\\point\\inward\\only} ([xshift=-5mm]dom.south west);
\end{tikzpicture}}
\caption{Architecture of APEX: the \apexlayers{} layers, from the front ends (top) to the configuration schemas (bottom); imports point inward only, as import-linter contracts check. Plugin packages register through entry points, and deck-based simulators run as external programs.}
\label{fig:architecture}
\end{figure}
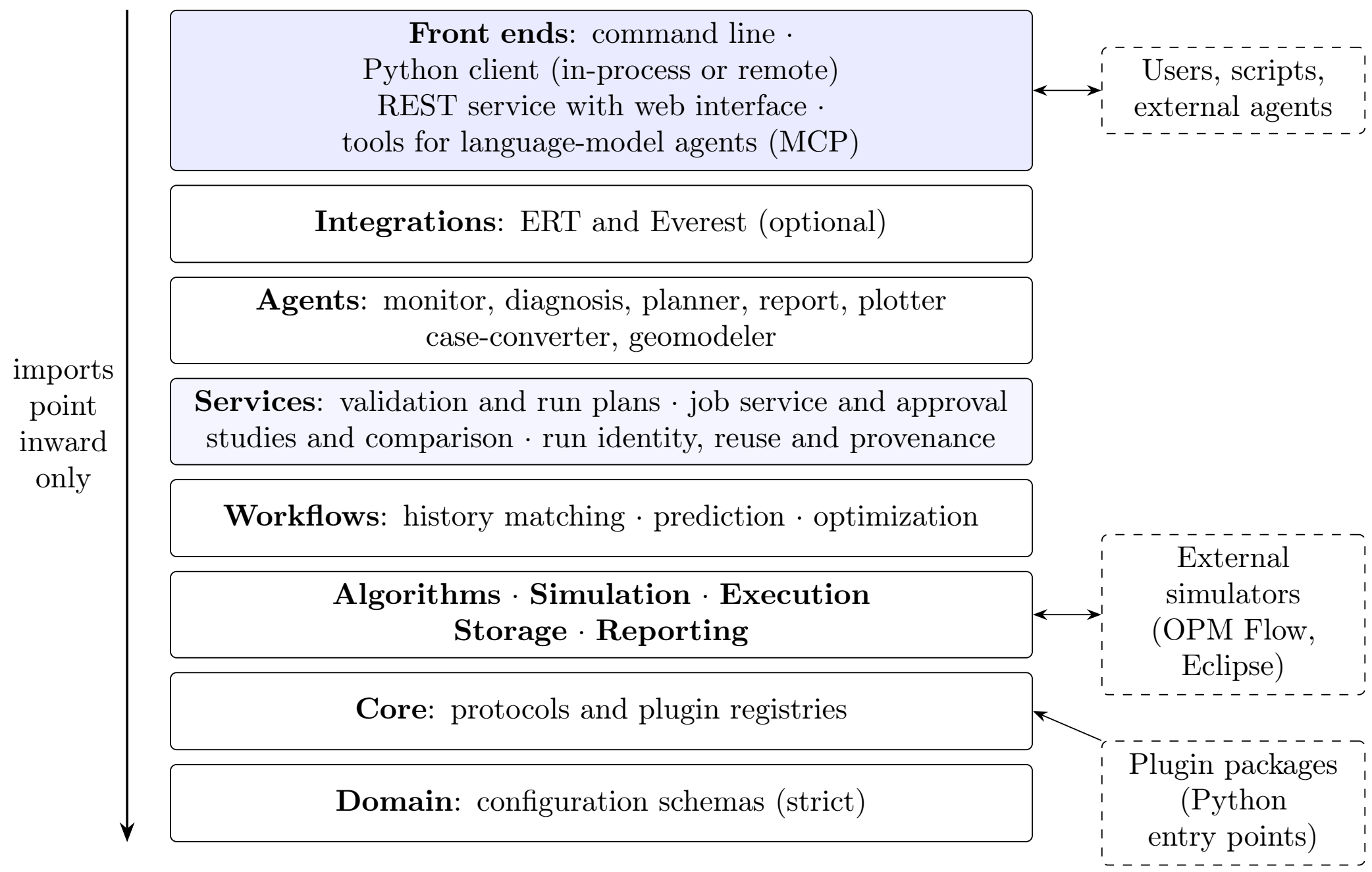

APEX is a Python package organized in \apexlayers{} layers (Figure~\ref{fig:architecture}). 
Inner layers never import outer ones; import-linter contracts check this rule, confine ERT and Everest to an integrations layer, and forbid any import of PET.
The innermost layer holds the configuration schemas as strict pydantic models, so that an unknown key is an error rather than a silent default. 
The core layer defines the protocols that plugins implement, for example, an ensemble algorithm, an update strategy, a simulator and an executor. 
The workflows layer has one generic driver per task (history matching, prediction, optimization): an algorithm owns its loop and requests forecasts, while the driver runs them, scores them and writes checkpoints. 
The services layer turns configurations into validated jobs, runs and compares studies; the agents layer reasons over the services' outputs and drafts cases and priors; and the outermost layer offers the front ends. 

Plugins are registered by name in \apexregistries{} plugin registries (Table~\ref{tab:plugins}): a run configuration selects from \apexconfigregistries{} of them, and the case-converter and the geomodeler use the other three, for their skills and the geomodeler's variogram estimator. 
Each plugin declares typed options, exposed as JSON Schemas, and may declare a status tag (\texttt{parity}, \texttt{validated}, \texttt{stable} or \texttt{experimental}). 
Other packages add plugins through Python entry points, so a new method or metric can be installed as a separate package without changing APEX. 

\subsection{Software functionalities}
\label{sec:functionalities}

\begin{table}[t]
\centering
\caption{Plugins of APEX by kind, with their declared status tags (``--'': none declared; on \texttt{parity}, see Section~\ref{sec:quality}). The list is not exhaustive; the reference pages in \texttt{docs/reference/plugins/} list every plugin.}
\label{tab:plugins}
\footnotesize
\begin{tabular}{@{}p{2.9cm}p{7.6cm}p{2.9cm}@{}}
\toprule
Kind & Plugins & Status tag \\
\midrule
History-matching methods & \texttt{es}, \texttt{esmda} \cite{emerick2012ensemble}, \texttt{gies} \cite{luo2021novel,luo2015Iterative}, \texttt{lm\_enrml}, \texttt{gn\_enrml} \cite{chen2013-levenberg,chen2012ensemble}, \texttt{multilevel\_esmda}; \texttt{chop} \cite{luo2022continuous,luo2023hyper}, with its components \texttt{data\_weights}, \texttt{gain\_inflation}, \texttt{length\_scale} & parity; \texttt{chop} and its components: -- \\
Update strategies & \texttt{approx}, \texttt{full}, \texttt{subspace}, \texttt{subspace2}, \texttt{rlmmac}, \texttt{gmac} \cite{luo2022data}, \texttt{local\_analysis}, others; \texttt{ert\_esmda}, \texttt{ert\_adaptive}, \texttt{ert\_localized} & --; \texttt{ert\_*}: validated \\
Localization & \texttt{adaptive}, \texttt{adaptive\_prior}, \texttt{adaptive\_per\_update}, \texttt{adaptive\_projected} \cite{luo2018correlation_spej,luo2019automatic}; \texttt{distance}, \texttt{distance\_entries} \cite{chen2010cross,Emerick2011combining} & -- \\
Priors, parameterizations & \texttt{gaussian}, \texttt{gaussian\_field}, \texttt{file}, \texttt{geostatspy\_sgsim}, \texttt{gstools\_pgs}, \texttt{mpslib\_snesim}; \texttt{facies\_distance}, \texttt{pgs\_truncation}; \texttt{control\_scaling}, \texttt{control\_schedule} & \texttt{control\_*}: stable; others -- \\
Simulators, executors & \texttt{opm\_flow} \cite{Rasmussen2020OPM}, \texttt{eclipse}, \texttt{ert}, \texttt{linear}, \texttt{toy\_reservoir}; \texttt{serial}, \texttt{process}, \texttt{ert} & \texttt{toy\_reservoir}: validated; others -- \\
Metrics, stopping, forecast scores & \texttt{rmse}, \texttt{ssim}, \texttt{normalized\_data\_mismatch}, facies scores; \texttt{patience}, \texttt{select} (cross validation \cite{luo2024better}); \texttt{crps}, \texttt{coverage\_p10\_p90}, others & -- \\
Geophysical data & AVO (petro-elastic models, reflectivity, attributes), gravity, seafloor displacement; \texttt{wavelet} compression & -- \\
Optimizers & \texttt{enopt}, \texttt{genopt}, \texttt{smcopt}, \texttt{line\_search}, \texttt{trust\_region}; \texttt{ga}, \texttt{pso}, \texttt{nsga2}, \texttt{nsga3}, \texttt{moead} (ported from pymoo \cite{blank2020pymoo}), \texttt{ropt} & parity; experimental \\
Response objectives & \texttt{npv}, \texttt{ren\_npv}, \texttt{ecalc\_npv}, \texttt{ecalc\_pareto\_npv}, \texttt{ren\_npv\_co2}; \texttt{field\_total}, \texttt{ecalc\_emissions} & parity; validated \\
Case-converter, geomodeler & converter skills \texttt{opm\_flow}, \texttt{eclipse}, \texttt{pet}, \texttt{ert}; geomodeling skills \texttt{continuous}, \texttt{correlated\_continuous}, \texttt{facies\_mps}, \texttt{facies\_pluri}; variogram estimator \texttt{wls} & -- \\
\bottomrule
\end{tabular}
\end{table}

\textit{History matching.}
Table~\ref{tab:plugins} lists several available ensemble-based methods, such as the ensemble smoother with multiple data assimilation (ES-MDA) and the generalized iterative ensemble smoother (GIES) with its regularized Levenberg--Marquardt (RLM-MAC) update. 
CHOP (Continuous Hyperparameter OPtimization) wraps an iterative method and estimates its hyper-parameters as it runs. 
Localization and damping schemes are auxiliary techniques that a compatible method may take; further options include soft constraints through a generalized update strategy, stopping by cross validation (CV) and an adjoint-based variant of the approximate update. 

\textit{Data, parameters and forecasts.}
Production data come from a table of observations, and field observations, such as amplitude versus offset (AVO) attributes, gravity and seafloor displacement, from the simulator's restart files, optionally compressed by wavelets \cite{luo2017ensemble}.
Prior ensembles are generated by geostatistical and/or facies simulations.
A finished history match can be forecast beyond history and scored, for instance by the continuous ranked probability score (CRPS).

\textit{Optimization.}
APEX ports PET's ensemble optimization algorithms without the need for derivatives, and line-search and trust-region methods for objectives with exact derivatives, all tagged \texttt{parity}.
Scaling and scheduling parameterizations map the controls, such as well pressures, onto the deck, and objectives such as the net present value (NPV) are computed from the simulator's responses.
Robust optimization evaluates each control vector on several model realizations, such as a history-matched ensemble.
Multi-objective optimizers ported from pymoo, tagged \texttt{experimental}, report their Pareto fronts with hyper-volumes.

\textit{Studies.}
A study expands cases $\times$ methods $\times$ variants $\times$ replicates into cells; a variant is a JSON merge patch over each case's configuration.
All variants share the replicates' seeds, so cells are paired by replicate.

\textit{Jobs and approval.}
Jobs run in background worker processes, scheduled by reservations against machine pools, such as simulator licences and processors.
A submission above the approval limits (by default \apexapprovalruns{} runs or \apexapprovalsimruns{} simulator runs), or whose cost cannot be bounded, needs approval; a grant approves in advance a budget of simulator runs for one client's experiment, but never a cost that cannot be bounded.
A policy can confine the files a job reads, executes and writes.

\textit{Front ends.}
The command line, the in-process and remote Python clients, the REST API, the web interface and the tools for language-model agents call the same service methods, so the same validation and approval limits apply to users, scripts and agents.
Who may approve differs: local command-line and in-process callers can confirm their own submissions, whereas REST and agent-tool clients need a token approved on a terminal (never over HTTP) or by a grant. 
These are implemented controls for clients that use APEX's interfaces, not a security boundary against a user with access to the services' files.

\textit{Agents.}
The runtime agents follow one design: code computes facts, named rules decide, and an optional language model's answer is held to the facts, so every agent also works without a model.
Among the currently available agents, five ones advise (monitor, diagnosis, planner, report, plotter), and two prepare cases: the case-converter can directly import an OPM Flow or Eclipse deck as an validated APEX case, or translate cases between APEX, PET and ERT; the geomodeler can generate a prior ensemble for a given case, optionally conditioned to specified hard data. 
No agent submits, stops or simulates without a user's confirmation, nor does the web interface's optional AI assistant.

\subsection{Software quality}
\label{sec:quality}
The code in APEX is type-checked by \texttt{mypy} in strict mode, linted by \texttt{ruff}, and its architectural dependencies are checked by \texttt{import-linter}. These checks are integrated into the repository's continuous integration (CI) workflow, which automatically verifies proposed code changes before they are merged. The CI workflow therefore provides an early check against type errors, style violations, architectural dependency violations, and regressions detected by the automated test suite.

The methods ported from third-party repositories, e.g., PET, are checked by parity tests (\texttt{tests/parity/}) against stored reference outputs that are regenerated in separate environments. 
These tests check consistency with the reference implementation at the level of individual update steps and rules, typically to $\apexparitytol$ times the step's largest entry, and replay whole optimizations, typically to a relative tolerance of $\apexpopttol$. 

For instance, \texttt{docs/migration/pet\_parity.md} gives each ported PET method status and tolerance, explains each divergence and lists the PET limits that APEX deliberately avoids. Meanwhile,
the optimizers ported from pymoo \cite{blank2020pymoo} are checked against its outputs: their operators draw pymoo's random numbers in pymoo's order and match exactly or to a relative tolerance of $\apexpymootol$, and the hypervolume to $\apexpymoohvtol$; complete runs of genetic algorithm (GA), particle swarm optimization (PSO), and non-dominated sorting genetic algorithm II (NSGA-II) are compared statistically, with \texttt{docs/migration \\ /pymoo.md} listing the existing differences. 

For the release described here, a run of the default suite (which excludes tests needing a real simulator) on a fresh clone, on Linux with Python \apexpython, gave \apextestspassed{} passed, \apextestsskipped{} skipped and \apextestsfailed{} failed tests\apextestsfailnote. 
The web interface's \apexuitests{} tests pass.
The code is type-checked by mypy in strict mode, linted by ruff, and its architecture is checked by import-linter. 

\section{Illustrative examples}
\label{sec:examples}

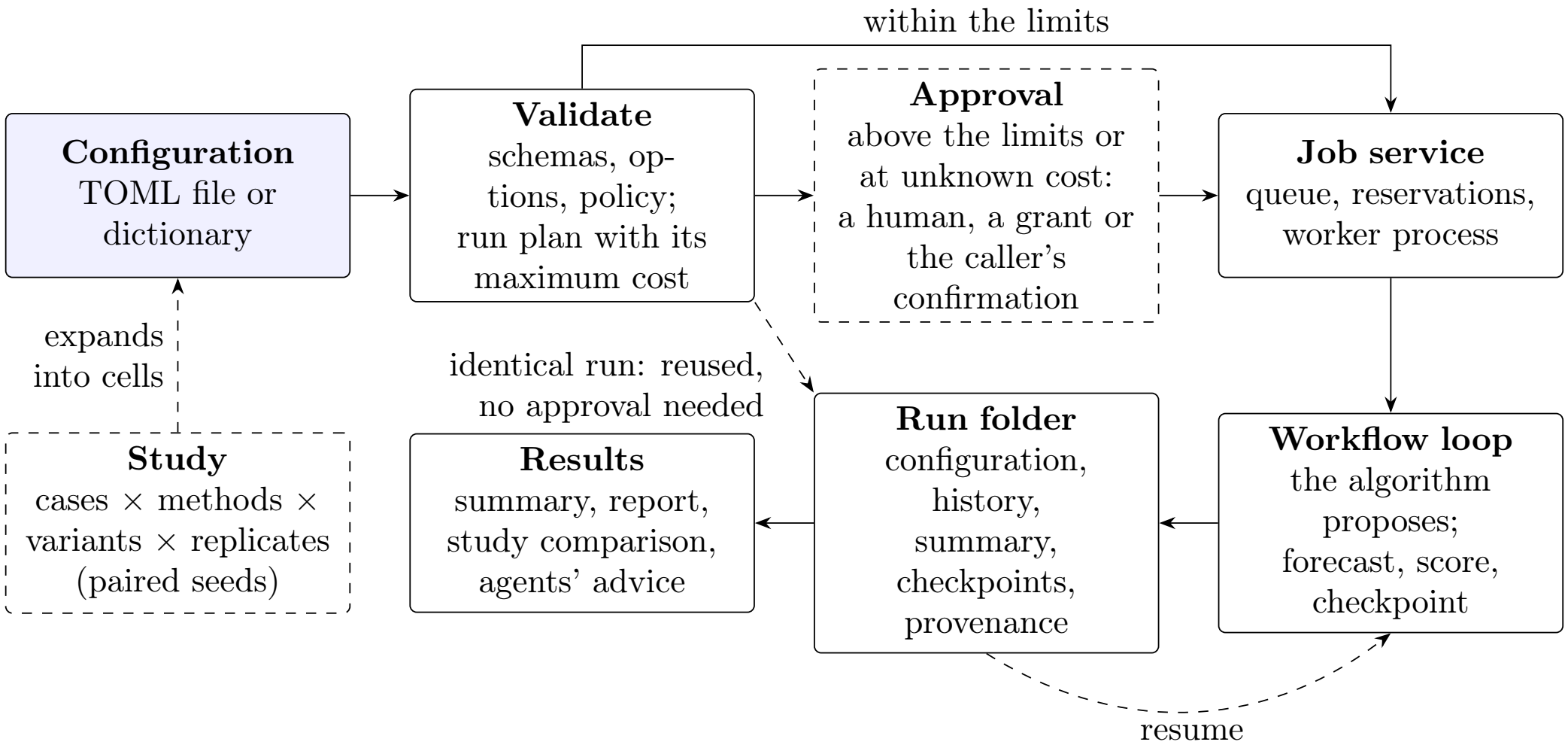
\begin{figure}[t]
\centering
\resizebox{\linewidth}{!}{%
\begin{tikzpicture}[
  wfstep/.style={draw, rounded corners=2pt, text width=3.3cm, minimum height=1.7cm,
    align=center, font=\footnotesize},
  >={Stealth}]
\node[wfstep, fill=blue!6] (cfg) at (0,0) {\textbf{Configuration}\\TOML file or dictionary};
\node[wfstep] (val) at (4.2,0) {\textbf{Validate}\\schemas, options, policy;\\run plan with its\\maximum cost};
\node[wfstep, dashed] (app) at (8.4,0) {\textbf{Approval}\\above the limits or\\at unknown cost:\\a human, a grant or\\the caller's confirmation};
\node[wfstep] (job) at (12.6,0) {\textbf{Job service}\\queue, reservations,\\worker process};
\node[wfstep] (loop) at (12.6,-3.4) {\textbf{Workflow loop}\\the algorithm proposes;\\forecast, score,\\checkpoint};
\node[wfstep] (run) at (8.4,-3.4) {\textbf{Run folder}\\configuration, history,\\summary, checkpoints,\\provenance};
\node[wfstep] (res) at (4.2,-3.4) {\textbf{Results}\\summary, report,\\study comparison,\\agents' advice};
\node[wfstep, dashed] (study) at (0,-3.4) {\textbf{Study}\\cases $\times$ methods $\times$\\variants $\times$ replicates\\(paired seeds)};
\draw[->] (cfg) -- (val);
\draw[->] (val) -- (app);
\draw[->] (app) -- (job);
\draw[->] (val.north) -- ++(0,0.45) -| node[pos=0.25, above, font=\footnotesize] {within the limits} (job.north);
\draw[->] (job) -- (loop);
\draw[->] (loop) -- (run);
\draw[->] (run) -- (res);
\draw[->, dashed] (study) -- node[left, font=\footnotesize, align=right] {expands\\into cells} (cfg);
\draw[->, dashed] (val.south east) -- node[pos=0.4, below left, font=\footnotesize, align=right] {identical run: reused,\\no approval needed} (run.north west);
\draw[->, dashed] (run.south) to[bend right=35] node[below, font=\footnotesize] {resume} (loop.south);
\end{tikzpicture}}
\caption{The workflow of a history-matching run in APEX. Approval (dashed) is needed only above the limits or when the cost cannot be bounded, and an identical configuration reuses its earlier run without approval.}
\label{fig:workflow}
\end{figure}

We illustrate the main functions of APEX on the repository's 3dBox example (\texttt{examples/\allowbreak 3dbox\_production\_avo/}), a synthetic three-phase reservoir of $60{\times}60{\times}5$ cells with six wells. 
ES-MDA or GIES, each with or without adaptive localization \cite{luo2018correlation_spej,luo2019automatic}, estimates the log-permeability and porosity of every cell (36,000 unknowns, 100 members) from the wells' monthly rates and root-mean-square maps of two amplitude-versus-offset (AVO) angle stacks at two survey dates, compressed by discrete wavelet transforms.
Each forecast chains OPM Flow \cite{Rasmussen2020OPM}, a petro-elastic model and an AVO model. 
Figure~\ref{fig:workflow} summarizes the workflow that the example follows. 

\textit{A configuration and its run plan.}
Listing~\ref{lst:config} is the main part of the example's ES-MDA configuration.
Each table names a plugin and its options, and \texttt{[run] seed} fixes the random numbers.
Validation (\texttt{apex validate}, Listing~\ref{lst:cli}) checks the file against the schemas and the plugins' option models, and returns a run plan without simulating anything, giving, among others, the ensemble size, the data per role and the maximum number of simulator runs.
For ES-MDA the maximum is exact, the ensemble size times one more than the number of inflation factors, here $100 \times (4+1) = 500$ OPM Flow runs; for an iterative method, such as GIES, it is the worst-case number that is used as the iteration cap. 
For Listing~\ref{lst:config}'s file, \texttt{apex validate} reports 100 members, 36,000 unknowns, 2,591 history-matching data, at most 5 forecasts and at most 500 simulator runs. 

\begin{lstlisting}[float=tp, caption={Excerpt of the example's ES-MDA configuration (\texttt{examples/\allowbreak 3dbox\_production\_avo/\allowbreak config\_esmda.toml}); comment lines, the state variables and other tables and lines are omitted.}, label=lst:config]
[run]
name = "3dbox_production_avo_esmda"
seed = 2026

[prior]   # ln PERMX and porosity correlation = 0.7
generator = { name = "file", options = { ensemble_file = "prior_ensemble.npz" } }

[observations]
file = "observations.csv"

[[observations.fields]]
time = "2027-06-24"
datatype = "avo_10"
file = "avo_10_2027-06-24.npy"
variance = 7.376470538467434e-07

[[data_transforms]]
name = "wavelet"
options = { datatype = "avo_10", wavelet = "db2", level = 3 }

[simulator]
name = "opm_flow"

[method]
name = "esmda"
options = { inflation = [4, 4, 4, 4] }
strategy = { name = "approx", options = { energy = 0.98 } }
localization = { name = "adaptive_prior", options = { threshold = "fixed" } }

[geophysics]
pem = { name = "standard" }

[[geophysics.data]]
kind = "avo"
datatype = "avo"
vintages = ["2027-06-24", "2032-12-14"]

[geophysics.data.options]   # the case's PET settings
angles = [10, 20]
reflectivity = "approx_zoeppritz_pp"
wavelet = { frequency = 45.0, length = 0.05 }                 # Ricker
\end{lstlisting}

\begin{lstlisting}[float=tp, language=bash, caption={Installing APEX, then validating and running the example and the study of Listing~\ref{lst:study} from the command line, as in the READMEs and the command reference.}, label=lst:cli]
git clone --branch v0.1.1 \
    https://github.com/lanhill/APEX-public.git apex && cd apex
uv sync --locked                       # locked environment
uv run apex validate examples/3dbox_production_avo/config_esmda.toml
uv run apex run examples/3dbox_production_avo/config_esmda.toml
uv run apex report \
    examples/3dbox_production_avo/runs/3dbox_production_avo_esmda
uv run apex study run examples/3dbox_production_avo/study.toml --wait
uv run apex study compare localization_avo
\end{lstlisting}

\textit{Running, the run folder and provenance.}
From Python, the class \texttt{apex.\allowbreak client.\allowbreak Client} offers the steps of Listing~\ref{lst:cli} as methods (\texttt{validate}, \texttt{submit\_\allowbreak run}, \texttt{wait}, \texttt{run\_\allowbreak summary}, \texttt{submit\_\allowbreak study}, \texttt{compare}), and \texttt{Client.connect(url)} returns a remote client with the same calls.
A run's folder holds the configuration as run, its history and summary, the event log, the ensembles in \texttt{checkpoints/}, and \texttt{provenance.json}, which records the APEX commit, the libraries, the host and the hashes of the inputs.
A run's identifier hashes its resolved configuration, with file paths replaced by their contents' hashes, together with the code version, so an identical configuration is reused instead of rerun.
Checkpoints include the random streams, so a stopped run resumes bit-identically on the same host, commit and processor class.
The run uses OPM Flow for forward simulations, with hours of computing for the 3dBox case on a workstation. 
It is also validated that two runs of Listing~\ref{lst:config}'s configuration on different machines produce identical results, at every iteration and in every metric.

\textit{A study and its paired comparison.}
For illustration, Listing~\ref{lst:study} compares ES-MDA without and with the example's localization over six paired replicates: 12 runs of at most $2 \times 6 \times 500 = 6{,}000$ OPM Flow runs, within the budget and the default approval limit.
Validation expands the study into its cells and plans each cell (case) individually; the budget refuses a study whose planned total number of forward simulations exceeds the limit, before anything runs.
For each case, metric and variant, \texttt{compare} gives the replicates' mean and standard deviation, as well as the paired mean difference, its standard error and the two-sided Wilcoxon signed-rank $p$-value \cite{wilcoxon1945individual} against the first variant.
Here, the replicates share the stored prior ensemble and differ only in their random perturbations, so the inference is conditional on the case and its prior.

\begin{lstlisting}[caption={A study of the example, which the reader saves as \texttt{examples/\allowbreak 3dbox\_production\_avo/\allowbreak study.toml} (not uploaded to the repository); its second variant's merge patch adds Listing~\ref{lst:config}'s localization to the case without it.}, label=lst:study]
[study]
name = "localization_avo"
seed = 11
replicates = 6
metrics = ["normalized_data_mismatch:all", "rmse:permx", "rmse:poro"]
budget = { max_simulator_runs = 6000 }

[study.cases]
avo = "config_esmda_noloc.toml"

[[study.variants]]
name = "none"

[[study.variants]]
name = "adaptive"
patch = { method = { localization = { name = "adaptive_prior", options = { threshold = "fixed" } } } }
\end{lstlisting}

\begin{figure}[!htb]
\centering
\includegraphics[width=\linewidth]{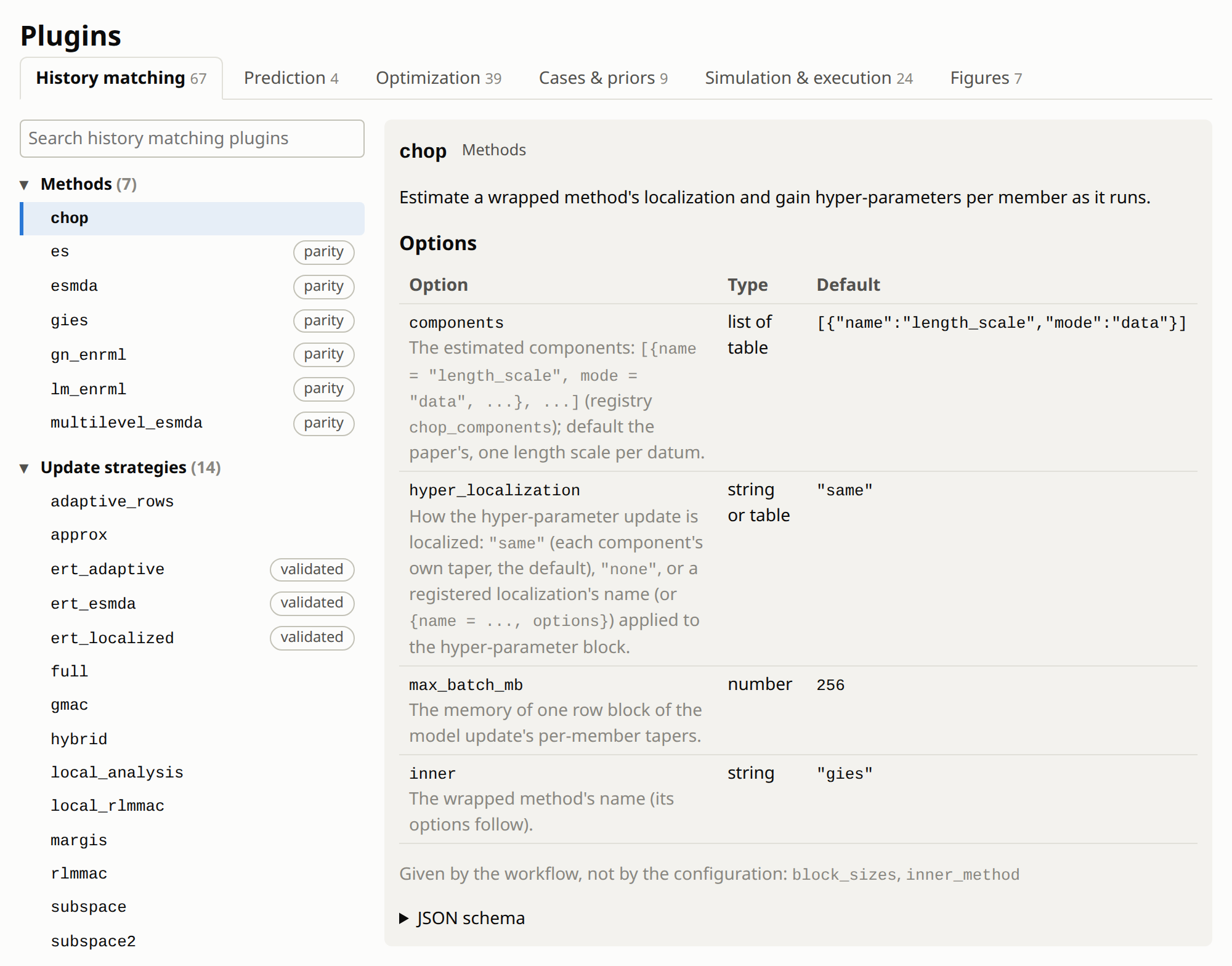}
\caption{The web interface's plugin catalogue, with the method \texttt{chop} selected (cropped). Each plugin is listed under its registry with its status tag; the panel shows the selected plugin's options, with their descriptions, types and defaults, generated from the code, and opens its JSON Schema.}
\label{fig:plugins}
\end{figure}

\begin{figure}[!htb]
\centering
\sbox{\shotbox}{\includegraphics{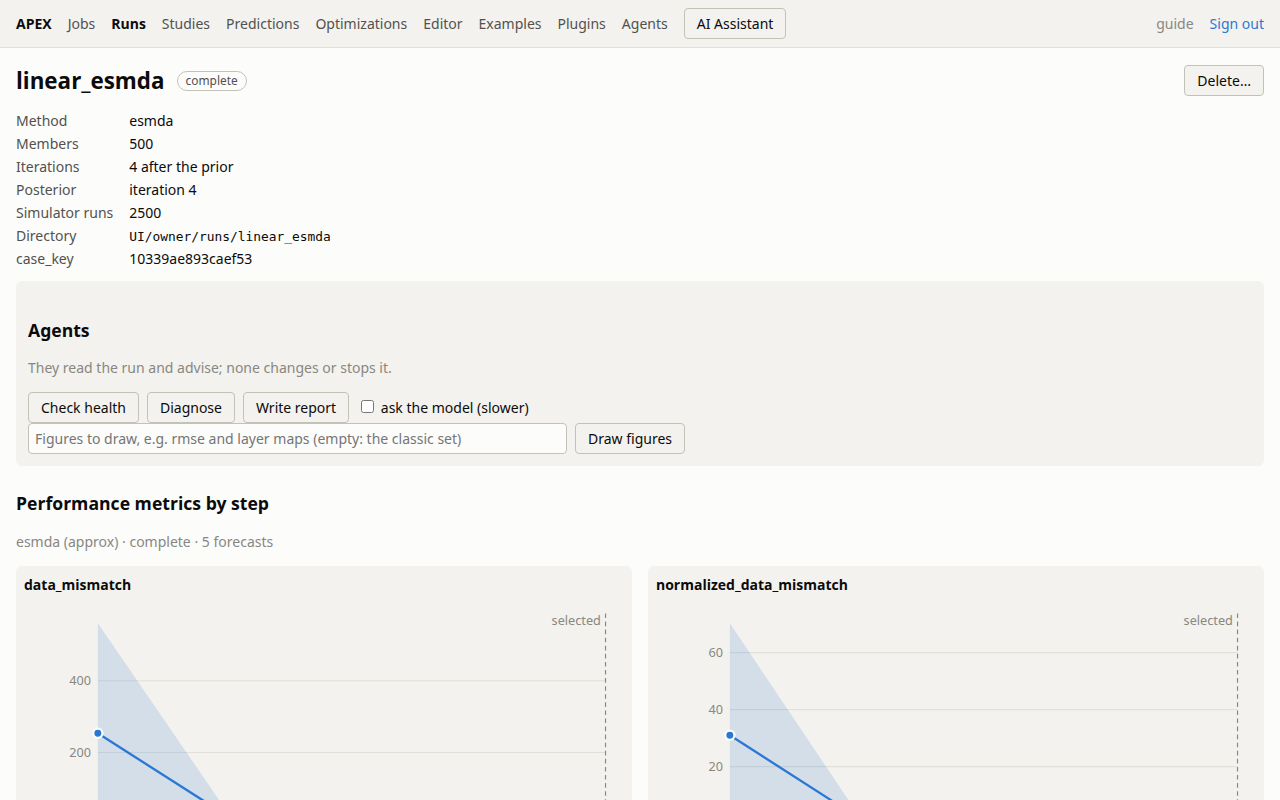}}%
\includegraphics[width=\linewidth, trim=0 {0.41\ht\shotbox} {0.375\wd\shotbox} 0, clip]{webui-run.png}
\caption{The header of a history-matching run's page in the web interface (cropped), for the documentation's linear-Gaussian tutorial run: the method, members, iterations, simulator runs and the buttons of the runtime agents, which read the run and advise. It illustrates the interface, not a result.}
\label{fig:run}
\end{figure}

\textit{Command line and web interface.}
\texttt{apex plugins} lists every plugin with its status tag, as the web interface does (Figure~\ref{fig:plugins}).
Figure~\ref{fig:run} shows the header of a run's page with the agents' buttons; the monitor, for instance, flags ensemble collapse when a variable's spread falls below \apexcollapse{} of the prior's.

\FloatBarrier

\section{Impact}
\label{sec:impact}

\textit{Comparisons with stated uncertainty, at a known cost.}
Studies are meant to make it routine to repeat a comparison over cases, variants and paired seeds, and to report a paired difference with its standard error and a test whose limits are stated, with each number traced to its configuration, inputs and code version.

Meanwhile, run plans, budgets, approval limits and grants make the maximum simulator cost explicit before submission, which matters when each forecast is an expensive reservoir simulation or licences are scarce.
A question such as whether an adaptive localization scheme improves a given method in each of several case studies, beyond seed-to-seed variability, can then be posed as a study file of a few lines.

\textit{Bridges to PET, ERT and one's own decks.}
PET (main branch commit e65fb10) and ERT (version 26.0.1) cases convert to and from APEX cases, within what each code can express (ERT's through the optional \texttt{ert} extra), and users of PET can rerun the parity tests.

With the same extra, APEX's history-matching methods run locally on an ERT case and write their iterates back to ERT's storage; conversely, a study can run ERT's ES-MDA or ensemble smoother on the same OPM Flow or Eclipse cases, seeds and priors as APEX's methods and compare them with the same paired statistics.

Methods from two open ecosystems can thus be compared within one study.
The case-converter and the geomodeler, which turn a deck with its observed history into a case with a generated prior ensemble, help lower the barrier to apply ensemble methods to one's own models.

\textit{New methods and extensions.}
APEX also adds methods without a PET or ERT counterpart, such as cross-validation based stopping criterion, CHOP, IES with soft constraints through a generalized model-update formula.

In APEX, a new method, metric or simulator can be installed as a package of its own, as the documentation demonstrates through a performance metric and a history-matching algorithm.

APEX's protocols are generic, and the linear and analytic simulators show that the workflows do not depend on a reservoir simulator. As such, it is expected that reuse of APEX beyond subsurface applications is plausible.
Additionally, external planners and language-model agents that use the REST API or the agent tools can drive experiments through the same validated configurations, and with the same budget limits as normal users.

\textit{Limitations.}
APEX is in early development (version \apexversion) and is tested on Linux with Python \apexpython{} only.
Its executors run locally or through ERT, without a scheduler yet for high-performance computing (HPC) clusters. Additionally, some plugins do not have status tags.
Furthermore, APEX is not on PyPI yet, while the web interface needs a separate build.

\section{Conclusions}
\label{sec:conclusions}

This work presents APEX, an open-source Python platform for ensemble-based history matching, prediction and optimization of subsurface models.
Its main idea is to treat an experiment, and a comparison of experiments, as a first-class object: a validated configuration with a planned maximum cost, a run with recorded provenance, and a study of paired replicates compared with paired statistics.
The checks of its implementation (e.g., through a continuous integration workflow) are part of the repository and can be rerun.
APEX is released under GPL-3.0-or-later at \url{https://github.com/lanhill/APEX-public}, as version \apexversion{} \cite{APEXSoftware}.






\section*{Acknowledgements}
XL acknowledges the financial support from the National Centre
for Sustainable Subsurface Utilization of the Norwegian Continental
Shelf (NCS2030), which is funded by industry partners and the Research
Council of Norway (project number: 331644).

\bibliographystyle{elsarticle-num}
\bibliography{references}

@misc{APEXSoftware,
  author       = {Xiaodong Luo},
  title        = {{APEX}: Agentic Platform for Ensemble eXploration},
  year         = {2026},
  howpublished = {Software, version 0.1.1 },
  note         = {\url{https://github.com/lanhill/APEX-public/tree/v0.1.1}},
  doi          = {10.5281/zenodo.23780462}
}

@article{BlankDeb2020Pymoo,
  author  = {Blank, Julian and Deb, Kalyanmoy},
  title   = {pymoo: Multi-Objective Optimization in Python},
  journal = {IEEE Access},
  year    = {2020},
  volume  = {8},
  pages   = {89497--89509},
  doi     = {10.1109/ACCESS.2020.2990567}
}

@INPROCEEDINGS{skjervheim2011ensemble,
  author = {Skjervheim, Jan-Arild and Evensen, Geir and Hove, J. and Vab{\o}, J. G.},
  title = {An ensemble smoother for assisted history matching},
  booktitle = {SPE Reservoir Simulation Symposium},
  year = {2011},
  address = {The Woodlands, Texas, USA},
  doi = {10.2118/141929-MS}
}

@ARTICLE{van1eeuwen996data,
  author = {Van Leeuwen, Peter Jan and Evensen, Geir},
  title = {Data assimilation and inverse methods in terms of a probabilistic
	formulation},
  journal = {Monthly Weather Review},
  year = {1996},
  volume = {124},
  pages = {2898--2913},
  doi = {10.1175/1520-0493(1996)124<2898:DAAIMI>2.0.CO;2}
}

@ARTICLE{chen2013-levenberg,
  author = {Chen, Y. and Oliver, D.},
  title = {{L}evenberg-{M}arquardt forms of the iterative ensemble smoother
	for efficient history matching and uncertainty quantification},
  journal = {Computational Geosciences},
  year = {2013},
  volume = {17},
  pages = {689--703},
  doi = {10.1007/s10596-013-9351-5}
}

@ARTICLE{emerick2012ensemble,
  author = {Emerick, Alexandre A and Reynolds, Albert C},
  title = {Ensemble smoother with multiple data assimilation},
  journal = {Computers \& Geosciences},
  year = {2013},
  volume = {55},
  pages = {3--15},
  publisher = {Elsevier},
  doi = {10.1016/j.cageo.2012.03.011}
}

@ARTICLE{luo2015Iterative,
  author = {Xiaodong Luo and Andreas Stordal and Rolf Lorentzen and Geir N{\ae}vdal},
  title = {Iterative ensemble smoother as an approximate solution to a regularized
	minimum-average-cost problem: theory and applications},
  journal = {SPE Journal},
  year = {2015},
  volume = {20},
  pages = {962--982},
  note = {{SPE-176023-PA}},
  doi = {10.2118/176023-PA}
}

@Article{luo2021novel,
  author  = {Luo, X},
  title   = {Novel iterative ensemble smoothers derived from a class of generalized cost functions},
  journal = {Computational Geosciences},
  year    = {2021},
  volume  = {25},
  pages   = {1159--1189},
  doi     = {10.1007/s10596-021-10046-1}
}

@article{luo2017ensemble,
  title={An ensemble 4D-seismic history-matching framework with sparse representation based on wavelet multiresolution analysis},
  author={Luo, Xiaodong and Bhakta, Tuhin and Jakobsen, Morten and N{\ae}vdal, Geir},
  journal={SPE Journal},
  volume={22},
  number={03},
  pages={985--1010},
  year={2017},
  publisher={SPE},
  doi={10.2118/180025-PA}
}

@article{evensen2019efficient,
  title={Efficient Implementation of an Iterative Ensemble Smoother for Data Assimilation and Reservoir History Matching},
  author={Evensen, Geir and Raanes, Patrick N and Stordal, Andreas S and Hove, Joakim},
  journal={Frontiers in Applied Mathematics and Statistics},
  volume={5},
  pages={47},
  year={2019},
  publisher={Frontiers},
  doi={10.3389/fams.2019.00047}
}

@BOOK{Oliver2008,
  title = {Inverse Theory for Petroleum Reservoir Characterization and History
	Matching},
  publisher = {Cambridge University Press},
  year = {2008},
  author = {Oliver, D.~S. and Reynolds, A.~C. and Liu, N.}
}

@ARTICLE{chen2012ensemble,
  author = {Chen, Yan and Oliver, Dean S},
  title = {Ensemble randomized maximum likelihood method as an iterative ensemble
	smoother},
  journal = {Mathematical Geosciences},
  year = {2012},
  volume = {44},
  pages = {1--26},
  publisher = {Springer},
  doi = {10.1007/s11004-011-9376-z}
}

@ARTICLE{chen2010cross,
  author = {Chen, Yan and Oliver, Dean S},
  title = {Cross-covariances and localization for {EnKF} in multiphase flow
	data assimilation},
  journal = {Computational Geosciences},
  year = {2010},
  volume = {14},
  pages = {579--601},
  publisher = {Springer},
  doi = {10.1007/s10596-009-9174-6}
}

@ARTICLE{Emerick2011combining,
  author = {Emerick, Alexandre and Reynolds, Albert},
  title = {Combining sensitivities and prior information for covariance localization
	in the ensemble {K}alman filter for petroleum reservoir applications},
  journal = {Computational Geosciences},
  year = {2011},
  volume = {15},
  pages = {251--269},
  publisher = {Springer},
  doi = {10.1007/s10596-010-9198-y}
}

@Article{luo2018correlation_spej,
  author  = {Luo, Xiaodong and Bhakta, Tuhin and N{\ae}vdal, Geir},
  title   = {Correlation-based adaptive localization with applications to ensemble-based {4D} seismic history matching},
  journal = {SPE Journal},
  year    = {2018},
  volume  = {23},
  pages   = {396--427},
  note    = {{SPE-185936-PA}},
  doi     = {10.2118/185936-PA},
}

@Article{luo2019automatic,
  author    = {Luo, Xiaodong and Bhakta, Tuhin},
  title     = {Automatic and adaptive localization for ensemble-based history matching},
  journal   = {Journal of Petroleum Science and Engineering},
  year      = {2020},
  volume    = {184},
  pages     = {106559},
  publisher = {Elsevier},
  doi = {10.1016/j.petrol.2019.106559},
}

@article{luo2022continuous,
  title     ={{Continuous Hyper-parameter OPtimization (CHOP)} in an ensemble {K}alman filter},
  author    ={Luo, Xiaodong and Xia, Chuan-An},
  journal   ={Frontiers in Applied Mathematics and Statistics},
  year      ={2022},
  volume    = {8},
  pages     = {1021551},
  publisher = {Frontiers},
  doi       = {10.3389/fams.2022.1021551}
}

@article{luo2023hyper,
  title={Hyper-parameter optimization for improving the performance of localization in an iterative ensemble smoother},
  author={Luo, Xiaodong and Cruz, William C and Zhang, Xin-Lei and Xiao, Heng},
  journal={Geoenergy Science and Engineering},
  volume={231},
  pages={212404},
  year={2023},
  publisher={Elsevier},
  doi={10.1016/j.geoen.2023.212404}
}

@article{luo2024better,
  title={Better stopping through cross validation in an iterative ensemble smoother: A perspective from supervised machine learning},
  author={Luo, Xiaodong and Ranazzi, Paulo},
  journal={Geoenergy Science and Engineering},
  volume={243},
  pages={213298},
  year={2024},
  publisher={Elsevier},
  doi={10.1016/j.geoen.2024.213298}
}

@Article{luo2022data,
  author    = {Luo, Xiaodong and Cruz, William C},
  title     = {Data assimilation with soft constraints ({DASC}) through a generalized iterative ensemble smoother},
  journal   = {Computational Geosciences},
  year      = {2022},
  volume    = {26},
  number    = {3},
  pages     = {571--594},
  publisher = {Springer},
  doi       = {10.1007/s10596-022-10137-7},
}

@article{Rasmussen2020OPM,
  title={The {Open Porous Media Flow} reservoir simulator},
  author={Rasmussen, A.F. and Sandve, T.H. and Bao, K. and Lauser, A. and Hove, J. and Skaflestad, B. and Kl{\"o}fkorn, R. and Blatt, M. and Rustad, A.B. and S{\ae}vareid, O. and Lie, K. and Thune, A.},
  journal={Computers and Mathematics with Applications},
  volume={81},
  pages={159--185},
  year={2021},
  publisher={Elsevier},
  doi={10.1016/j.camwa.2020.05.014}
}

@misc{PETSoftware,
  author       = {{NORCE Norwegian Research Centre, PET group}},
  title        = {{PET}: Python Ensemble Toolbox},
  year         = {2026},
  howpublished = {Software, commit e65fb10 (GPL-3.0)},
  url          = {https://github.com/Python-Ensemble-Toolbox/PET}
}

@misc{ERTSoftware,
  author       = {{Equinor ASA}},
  title        = {{ERT}: Ensemble based Reservoir Tool},
  year         = {2026},
  howpublished = {Software, version 26.0.1 (GPL-3.0-only)},
  url          = {https://github.com/equinor/ert}
}

@inproceedings{Hanea2025Everest,
  author    = {Hanea, R. G. and Oliveira, R. and Popa, T. and Feldmann, F. and Eide, {\O}. and Sortland, S. and van der Heijden, T. L. M. and Barros, E. G. D. and Szklarz, S. P. and Leeuwenburgh, O. and Verveer, P. J. and de Hoop, S. and Bottero, S. and Hopstaken, K. M.},
  title     = {{Everest\texttrademark}: An Open-Source Platform for Decision Making Under Uncertainties},
  booktitle = {SPE Reservoir Simulation Conference},
  address   = {Galveston, Texas, USA},
  year      = {2025},
  doi       = {10.2118/223850-MS}
}

@techreport{White2020PESTpp,
  author      = {White, Jeremy T. and Hunt, Randall J. and Fienen, Michael N. and Doherty, John E.},
  title       = {Approaches to highly parameterized inversion: {PEST++} Version 5, a software suite for parameter estimation, uncertainty analysis, management optimization and sensitivity analysis},
  institution = {U.S. Geological Survey},
  type        = {Techniques and Methods},
  number      = {7-C26},
  year        = {2020},
  doi         = {10.3133/tm7C26}
}

@article{Raanes2024DAPPER,
  author  = {Raanes, Patrick N. and Chen, Yumeng and Grudzien, Colin},
  title   = {{DAPPER}: Data Assimilation with {Python}: a Package for Experimental Research},
  journal = {Journal of Open Source Software},
  volume  = {9},
  number  = {94},
  pages   = {5150},
  year    = {2024},
  doi     = {10.21105/joss.05150}
}

@inproceedings{Greff2017Sacred,
  author    = {Greff, Klaus and Klein, Aaron and Chovanec, Martin and Hutter, Frank and Schmidhuber, J{\"u}rgen},
  title     = {The {Sacred} Infrastructure for Computational Research},
  booktitle = {Proceedings of the 16th Python in Science Conference},
  pages     = {49--56},
  year      = {2017},
  doi       = {10.25080/shinma-7f4c6e7-008}
}

@article{blank2020pymoo,
  author  = {Blank, J. and Deb, K.},
  title   = {pymoo: Multi-objective optimization in {Python}},
  journal = {IEEE Access},
  volume  = {8},
  pages   = {89497--89509},
  year    = {2020},
  doi     = {10.1109/ACCESS.2020.2990567}
}

@article{wilcoxon1945individual,
  author  = {Wilcoxon, F.},
  title   = {Individual comparisons by ranking methods},
  journal = {Biometrics Bulletin},
  volume  = {1},
  number  = {6},
  pages   = {80--83},
  year    = {1945},
  doi     = {10.2307/3001968}
}

\end{document}